\documentclass[aps,prl,twocolumn,amsmath,showpacs,letterpaper]{revtex4}
\usepackage[utf8]{inputenc}   
\usepackage[T1]{fontenc}        
\usepackage{graphicx}           
\usepackage[plain]{fancyref}    
\usepackage{hyperref}

\graphicspath{{images/}}         

\newcommand{\bra}[1]{\left\langle #1\right|}
\newcommand{\ket}[1]{\left| #1\right\rangle}
\newcommand{\braket}[2]{\left\langle #1|#2\right\rangle}
\newcommand{\ketbra}[3][]{\left|#2\right\rangle_{#1}\!\left\langle#3\right|}

\newcommand{\sub}[1]{\textrm{\scriptsize {#1}}}  

\begin{document}
\title{Raman Adiabatic Transfer of Optical States}

\author{Jürgen Appel, K.-P. Marzlin, A. I. Lvovsky}

\affiliation{Institute for Quantum Information Science, University of
  Calgary, Calgary, Alberta T2N 1N4,
  Canada\footnote{\url{http://www.iqis.org/}}}

\date{\today}

\begin{abstract}
  We analyze electromagnetically induced transparency and light
  storage in an ensemble of atoms with multiple excited levels
  (multi-$\Lambda$ configuration) which are coupled to one of the
  ground states by quantized signal fields and to the other one via
  classical control fields. We present a basis transformation of
  atomic and optical states which reduces the analysis of the system
  to that of EIT in a regular 3-level configuration. We demonstrate
  the existence of dark state polaritons and propose a protocol to
  transfer quantum information from one optical mode to another by an
  adiabatic control of the control fields.
\end{abstract}


\maketitle

\section{Introduction} \label{sec:Introduction}

Electromagnetically induced transparency (EIT) is a quantum
interference effect occurring when a weak signal light field and a
stronger control field both interact with an ensemble of atoms with
$\Lambda$-shaped energy level configuration
\cite{harris97:_elect,arimondo96:_coher}. The quantum probabilities
for an excitation of the atoms by both light fields interfere
destructively, so that no excitation takes place and the normally
highly opaque medium becomes transparent for the signal field.

EIT in atomic media attracts great interest due to its possible
applications in nonlinear optics and quantum information processing.
In particular, high sensitivity to the two-photon resonance condition
leads to a steep dispersion for the signal field which therefore
experiences a greatly reduced group velocity. The demonstration of
such an effect in an ultra-cold atomic
gas~\cite{hau99:_light_speed_reduction} and hot atomic
vapor~\cite{scully99:_ultras_group_veloc_enhan_nonlin} and the
subsequent stopping of
light~\cite{hau01:_coherent_optical_information_storage,lukin01:_storag_light_atomic_vapor}
by an adiabatic process make this system appealing as a candidate for
a quantum optical memory device.

Of further interest are double- and multi-$\Lambda$ configurations
that contain two or more excited levels and are excited by several
control fields.  Nonlinear effects such as four-wave mixing
\cite{deng03:_inhib,merriam00:_effic_lambd}, phase conjugation
\cite{hemmer96:_optic_raman}, and amplification without inversion
\cite{kocharovskaya90:_amplif,Scully90_LukinRev} have been
investigated for strong fields applied to both sides of the~$\Lambda$
\cite{lukin00:_advan_atomic_molec_optic_physic}.  If, on the other
hand, the control fields couple to the same ground state, and the
signal fields to the other (\fref{fig:multilambda1}), the behavior of
the system with respect to the signal fields is analogous to regular
EIT, but with given control fields EIT is experienced by only one
specific linear combination of signal modes
\cite{cerboneschi95:_trans_lambd,cerboneschi96:_propag_lambd,korsunsky99:_phase,liu04:_dynam_symmet_its_applic_in,lukin00:_advan_atomic_molec_optic_physic}
whereas others get absorbed.  The action of the atomic sample on the
signal fields is analogous to that of an interferometer followed by
absorption of all but one output modes.  Raczy\'{n}ski and Zaremba
\cite{raczynski02:_contr,raczynski04:_elect} investigated formation of
dark-state polaritons
\cite{fleischhauer02:_quantum_memory_for_photons} as well as storage
of light in a double-$\Lambda$ system.

Most of the existing work on EIT in multilevel systems was done with
classical fields. An expansion into the quantum domain was undertaken
by Liu {\it et al.} who derived an expression for a dark state with
multiple quantum signal fields in stationary
modes~\cite{liu04:_dynam_symmet_its_applic_in,li04:_manip_synch_optic_signal_doubl}.
However, to our knowledge, no full quantum EIT/light storage formalism
has been developed for propagating optical fields in this system. In
the present paper, we bridge this gap by elaborating a basis
transformation for both atomic and optical states which reduces
multi-level EIT to the well investigated EIT in a regular $\Lambda$
scheme.  In addition, we show that by an adiabatic change of the
control fields, a transfer of quantum optical states between different
signal modes or their linear combinations can be implemented. This
procedure resembles stimulated Raman adiabatic passage
(STIRAP)~\cite{bergmann98:_stirap}, but applies to optical rather than
atomic states and can be useful for routing and distribution of
optically encoded information in classical and quantum communication.

\section{Discrete Field Modes} \label{sec:Discrete_Mode_Field}

\begin{figure}[htbp]
  \centering
  \includegraphics[width=0.6\columnwidth,keepaspectratio]{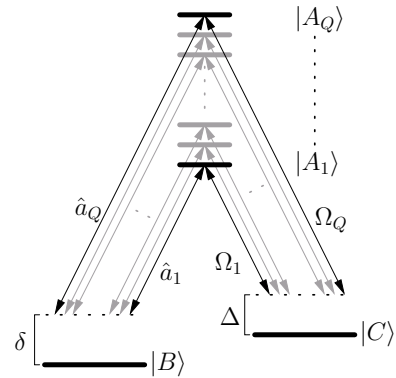}
  \caption{Multi $\Lambda$-system: $Q$ excited states $\ket{A_q}$ are each coupled
    by a classical control field $\Omega_q$ with detuning $\Delta$ to
    the ground state $\ket{C}$ and by a quantized field $\hat a_q$
    with detuning $\delta$ to another ground state $\ket{B}$.}
  \label{fig:multilambda1}
\end{figure}

In order to better understand EIT in a multi-$\Lambda$ ensemble, we
first focus on a simplified system with discrete (non-propagating,
such as in a cavity) quantized field modes before we generalize our
treatment to propagating wave packets.

We consider an ensemble of $N$ multi-$\Lambda$-configuration atoms
(\fref{fig:multilambda1}).  Each of the excited states
$\left\{\ket{A_{q}}\right\}_{q=1,\dotsc, Q}$ is coupled to the two
ground states $\ket{B}$ and $\ket{C}$ by a quantized signal field
$\hat a_{q}$ and a classical control field~$\Omega_{q}$, respectively.
All the signal beams are detuned from the optical resonance by the
same amount $\delta=E_q-E_B-\hbar \nu_q$; the detuning of the control
beams is $\Delta=E_q-E_C-\hbar \omega_q$, where $\nu_q$ and $\omega_q$
are the respective laser frequencies.

Let $\hat \sigma^j_{\alpha \beta}=\ketbra[jj]{\alpha}{\beta}$ be the
flip operator between the states $\ket{\beta}$ and $\ket{\alpha}$ of
the $j$-th atom. When all fields are resonant ($\delta=\Delta=0$) the
dynamics of this system is described by the interaction Hamiltonian
\begin{equation}
  \label{eq:1}
    \hat H_\sub{int}  = -\hbar \sum \limits_{j=1}^N  \sum \limits_{q=1}^Q \left( g_q \hat a_q \, \hat \sigma^j_{a_qb} + \Omega_q(t) \, \hat \sigma^j_{a_qc} \right)+  \text{H.a.}
\end{equation}
in the co-rotating frame.  Here $\hat a_q$ is the photon annihilation
operator of the $q$-th mode and $g_q$ describes the \mbox{vacuum} Rabi
frequency of that transition which is assumed to be equal for all the
atoms (Dicke limit). $\Omega_q(t)$~is the slowly varying Rabi
frequency of the according classical control field.

Let $\ket{\mathbf C^k}$ denote the totally symmetric state with $k$
atoms in state $\ket{C}$ and all others in state $\ket{B}$.
\begin{equation}
 \label{eq:2}
  \ket{\mathbf C^k}  =  \frac{1}{\sqrt{\binom{N}{k}}} \sum \limits_{1 \le j_1 < \dotsb < j_k \le N} \!\!\!\!\!\! \ket{B_1,\dotsc, C_{j_1},\dotsc, C_{j_k},\dotsc, B_N}
\end{equation}

By analogy to ref.~\cite{li04:_manip_synch_optic_signal_doubl}, it
then can be shown that the states
\begin{multline}
  \label{eq:3}
  \!\!\!\!\ket{D,n} \! = \! \left[ \sum \limits_{j=1}^{N} \hat
    \sigma^j_{CB} -\sum \limits_{q=1}^{Q} \left(\frac{\Omega_q}{g_q}
      {\hat a^\dagger}_q\right)\right]^n
  \!\!\!\ket{\mathbf{C}^0,(0,\dotsc,0)}
\end{multline}
are dark states: they are eigenstates of the interaction Hamiltonian
with zero eigenvalue. Here the $(n_1,\ldots,n_q)$-part denotes the
state of the quantized light field in Fock representation.

\subsection{Adiabatic transfer of optical states}

If one of the control fields is strong ($\Omega_i\gg g_i\sqrt{N})$)
while others vanish, the dark state takes the form
\begin{equation}
  \label{eq:4}
  \ket{D,n} \xrightarrow{\Omega_{k\ne i}\to0} \underset{\quad{i\textrm{th mode}}}{\ket{\mathbf C^0,(0,\dotsc,n,\dotsc,0)}};
\end{equation}
all photons gather in the according signal field mode.  If all
controls are slowly switched off the dark state adiabatically changes
to
\begin{equation}
  \label{eq:5}
  \ket{D,n} \xrightarrow{\Omega_1=\dotsb=\Omega_Q=0} \ket{\mathbf C^n,(0,\dotsc,0)},
\end{equation} so the quantum optical state carried by the $i$th mode is converted to a coherent collective ground state
superposition \cite{fleischhauer02:_quantum_memory_for_photons}.

Suppose now that while the system is in the state (\ref{eq:4}),
another control field $\Omega_j$ is turned on. In this case, by
adiabatic following, the state of the system will convert to
\begin{equation} \label{interm}
\ket{D,n}\xrightarrow{\Omega_{k\ne i,k\ne j}=0} \left(\frac{\Omega_i}{g_i} {\hat a^\dagger}_i+\frac{\Omega_j}{g_j} {\hat a^\dagger}_j\right)^n\ket{\mathbf C^0,(0,\dotsc,0)}.
\end{equation}
If $\Omega_i$ is then slowly turned off, the quantum state of the
$i$th optical mode will be transferred to the $j$th mode completely:
\begin{equation}
  \label{eq:7}
  \ket{D,n} \xrightarrow{\Omega_{k\ne j}\to 0} \underset{\quad{j\textrm{th mode}}}{\ket{\mathbf C^0,(0,\dotsc,n,\dotsc,0)}}.
\end{equation}
We see that, by varying the control fields, the quantum state can be
transferred to any other optical mode or their coherent superposition.
We call this procedure \emph{Raman adiabatic transfer of optical
  states} (RATOS) by analogy to the well known STIRAP technique which
permits transfer of population between atomic states by means of
adiabatic interaction with light \cite{bergmann98:_stirap}.  In RATOS,
on the other hand, quantum states are transferred between optical
states by adiabatic interaction with atoms.

The above treatment is valid for the case of discrete, non-propagating
modes, e.g. in a cavity. In the practical case of a propagating field,
photons first travel through an atom-free environment, then couple
into an EIT medium, experience RATOS while in transfer, and finally
leave the medium. In order to understand the propagation dynamics, the
theory must be reformulated in terms of dark-state polaritons akin to
ref.~\cite{fleischhauer02:_quantum_memory_for_photons}. This is our
task for the remainder of the paper.

One specific question that needs to be addressed is whether RATOS can
be applied to optical fields that are initially (prior to coupling
into an EIT system) not in a dark state in the sense of \fref{eq:3}.
An example is both modes $i$ and $j$ containing one photon while the
remaining modes are in the vacuum state. Can one choose control fields
in such a way that these photons are losslessly coupled into an EIT
medium, and if not, what minimum loss can one expect? More generally,
what are the possibilities of quantum optical state engineering in a
multi-level EIT environment?

\section{Mapping to a single-\texorpdfstring{$\Lambda$}{Lambda} system}
Our approach is to develop a basis transformation of the atomic and
optical states that will map a multi-$\Lambda$-system to a normal EIT
(single-$\Lambda$) scheme, thus providing an intuitive understanding
for the optical properties of the system.
\subsection{Changing the atomic basis}
Consider one atom with a multi-$\Lambda$ level structure as in
\fref{fig:multilambda1}. In the rotating wave frame the Hamiltonian
reads
\begin{equation}
  \label{eq:10}
  \begin{split}
    \hat H(t)
    & = -\hbar \frac{\delta}{2} \ketbra{B}{B} - \hbar \frac{\Delta}{2} \ketbra{C}{C} \\
    & \quad -\hbar \sum \limits_{q=1}^{Q} \Bigl( g_q \hat a_q
    \ketbra{A_q}{B} + \Omega_q(t) \ketbra{A_q}{C} \Bigr) \enspace +
    \text{H.a.},
  \end{split}
\end{equation}
which in the absence of the quantized fields reduces to
\begin{equation}
  \label{eq:11}
  \hat H_0 =-\hbar \frac{\delta}{2} \ketbra{B}{B}  -\hbar \left(\frac{\Delta}{2} \ket{C} +  \Omega \ket{EB}\right)\bra{C}  \quad +\text{H.a.,}\\
\end{equation} where
\begin{equation}
  \label{eq:12}
  \ket{EB} =  \sum \limits_{q=1}^Q \frac{\Omega_q}{\Omega} \ket{A_q} \text{ and}\\
\end{equation} and
\begin{equation}
  \label{eq:13}
  \Omega = \sqrt{\sum \limits_{q=1}^Q |\Omega_q|^2}.
\end{equation}
$\hat H_0$ possesses $Q$ eigenstates of zero eigenvalue, one of them
obviously being $\ket{B}$. The others are superpositions of excited
states $\ket{A_q}$ that are orthogonal to the ``excited bright state''
$\ket{EB}$ and thus not coupled by the control fields.

A basis spanning this subspace can be explicitly constructed by an
unitary Householder reflection \cite{lehoucq96:_comput}
\begin{gather}
  \label{eq:15}
  \hat U= \sigma \ketbra{u}{u}-\mathbf{1}
  \text{ with } \sigma=\braket{A_Q}{EB}+1=\frac{\Omega_Q}{\Omega}+1 \\
  \text{and } \ket{u}=\tfrac{1}{\sigma} (\ket{A_Q}+\ket{EB}) \nonumber
\end{gather}
so that $\ket{EB} = \hat U \ket{A_Q}$ and
\begin{gather}
  \label{eq:16}
  \begin{split}
    \ket{ED_q} \equiv \hat U \ket{A_q} = \frac{\Omega_q^*}{\Omega_Q^*+
      \Omega} \Bigl( \ket{A_Q} + \ket{EB} \Bigr) - \ket{A_q} \\
    \text{for } q=1,\dotsc,Q-1.
  \end{split}
\end{gather}

In this basis the interaction Hamiltonian then reads
\begin{gather}
  \label{eq:17}
  \begin{split}
    \hat H
    & =      - \hbar \, \frac{\Delta}{2} \ketbra{C}{C}  - \hbar \, \frac{\delta}{2} \ketbra{B}{B} - \hbar \, \Omega \ketbra{EB}{C} \\
    & \quad -\hbar \sum \limits_{q=1}^{Q} \sum \limits_{r=1}^{Q-1}
    g_q \braket{ED_{r}}{A_q} \hat a_q \ketbra{ED_{r}}{B} \\
    & \quad -\hbar \sum \limits_{q=1}^{Q} g_q \braket{EB}{A_q} \hat
    a_q \ketbra{EB}{B} \quad + \text{H.a.}
  \end{split}
\end{gather}
As expected, the states $\ket{ED_q}$ do not undergo any interaction
with the classical fields $\Omega_q$ at all.

This can be interpreted physically by understanding that the phases
and amplitudes of the excited states $\ket{A_q}$ are such that the
probability amplitudes for a transition from the states $\ket{ED_q}$
to $\ket{C}$ interfere destructively, akin to dark states in a normal
EIT scheme --- hence we call the $\ket{ED_q}$ ``excited dark states''.

Also in close analogy to EIT, the ground state $\ket{C}$ is coupled to
only one particular superposition $\ket{EB}$ of the excited states
(the ``excited bright state''), where the transition probabilities
interfere constructively.

\begin{figure}[tbp]
  \centering
  \includegraphics[width=0.6\columnwidth,keepaspectratio]{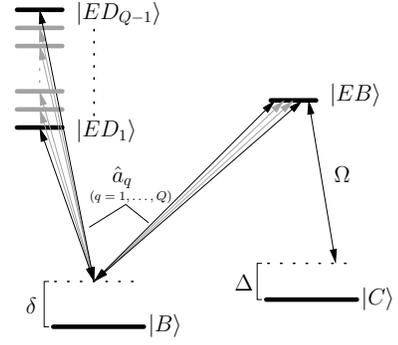}
  \caption{Multi $\Lambda$-system in the ``excited dark-state'' basis: The classic fields $\Omega_q$
    drive only the $\ket{EB} \Leftrightarrow \ket{C}$ transition and
    each quantized field mode $\hat a_q$ mode couples to all of the
    excited states $\ket{ED_{(1,\dotsc,Q-1)}}$ and $\ket{EB}$.}
  \label{fig:multilambda2}
\end{figure}

However, each of the weak quantized optical modes $\hat a_q$ couples
the ground state $\ket{B}$ to all of the states $\ket{EB},
\ket{ED_1},\dotsc,\ket{ED_{Q-1}}$ (see~\fref{fig:multilambda2}).  If
the excited states $\ket{A_q}$ have a short lifetime, so do the states
$\ket{ED_q}$. Hence, in general, light in the modes $\hat a_q$ would
not experience EIT; the photons would get absorbed, exciting the atom
to the $\ket{ED_q}$ levels which decay due to spontaneous emission. In
the next subsection we show, however, that there exists a linear
superposition of signal states which does not couple to
$\ket{ED_q}$'s, thus enabling EIT in this system.

\subsection{Changing the optical basis}

We are looking for a new set of quantized field mode operators $\hat
b_q$ defined by the unitary transform $\hat W$: $\hat a_q=\sum_{s=1}^Q
W_{qs} \, \hat b_{s}$, so that one of the new operators $\hat b_Q$
does not couple to any of the ``excited dark-states'' $\ket{ED_q}$, in
other words
\begin{equation}
  \label{eq:18}
  \sum \limits_{q=1}^{Q} g_q \braket{ED_r}{A_q} W_{qQ} \,\, \hat b_Q= 0 \quad \text{for all } r \neq Q.
\end{equation}

Since $\sum_{q=1}^{Q} \Omega_q \braket{ED_r}{A_q}=\braket{ED_r}{EB}=0$
we can choose
\begin{equation}
  \label{eq:19}
 W_{qQ} =\frac{1}{R} \frac{\Omega_q}{g_q} \quad  \text{ with } R = \sqrt{\sum_{q=1}^{Q} \left|\frac{\Omega_q}{g_q}\right|^2}
\end{equation}
as a solution for \fref{eq:18} and fix the other components of $\hat
W$ by constructing it as a Householder reflection in a fashion
analogous to \fref{eq:15}:
\begin{gather}
  \label{eq:20}
  W = \gamma \vec w \vec w^\dagger -\mathbf{1}
  \text{ with } \gamma= \frac{1}{R} \frac{\Omega_Q}{g_Q} +1 \\
  \text{and } \vec w = \frac{1}{\gamma} \left(\vec e_Q + \frac{1}{R}
    \sum \limits_{q=1}^{Q} \frac{\Omega_q}{g_q} \vec e_q\right)
  \nonumber.
\end{gather}

In this new atomic and photonic basis the Hamiltonian reads
\begin{gather}
  \label{eq:21}
  \begin{split}
    \hat H & = - \hbar \left(
      \frac{\Delta}{2} \ketbra{C}{C} + \Omega \ketbra{EB}{C} \right) \\
    & \quad - \hbar \left(\frac{\delta}{2} \ketbra{B}{B} + g \, \hat b_Q \ketbra{EB}{B} \right) \\
    & \quad -\hbar \sum \limits_{q=1}^{Q-1} \hat b_q \left( g_q^{EB}
      \ketbra{EB}{B}
      + \sum  \limits_{r=1}^{Q-1} g_q^{ED_r} \ketbra{ED_r}{B} \right) \\
    & \quad +\text{H.a.}
  \end{split}
\end{gather}

The first two terms correspond exactly to the Hamiltonian of a
traditional \mbox{$\Lambda$-system} $(\ket{B}\leftrightarrow \ket{EB}
\Leftrightarrow \ket{C})$. The quantized field mode
\begin{equation}
  \label{eq:22}
  \hat b_Q = \frac{1}{R} \sum \limits_{q=1}^Q \frac{{\Omega_q}^*}{{g_q}^*} \hat a_q
\end{equation}
couples $\ket{EB}$ to $\ket{B}$ with strength $g=\tfrac{\Omega}{R}$
and detuning $\delta$ whereas, among all excited atomic states, only
$\ket{EB}$ is coupled to $\ket{C}$ by the classical field
mode~$\Omega$ detuned by~$\Delta$.

We see that weak signal pulses in the $\hat b_Q$ mode interact with
the atoms of a multi-\mbox{$\Lambda$-medium} in a fashion completely
analogous to pulses propagating through the well understood standard
EIT system.

\begin{figure}[tbp]
  \centering
  \includegraphics[width=0.6\columnwidth,keepaspectratio]{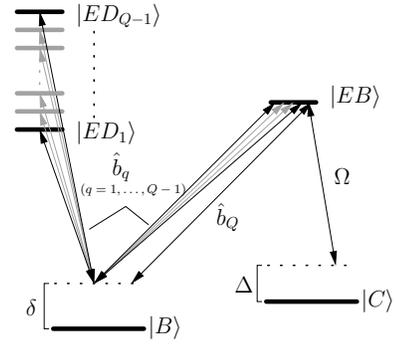}
  \caption{The multi $\Lambda$-system after basis transformation 
    of both atomic and optical Hilbert spaces: The classical
    fields~$\Omega_q$ drive only the $\ket{EB} \Leftrightarrow
    \ket{C}$ transition and form an EIT system with the quantized
    field mode $\hat b_Q$.  The other modes $\hat b_{q\neq Q}$ couple
    also to the excited dark states $\ket{ED_q}$ and therefore undergo
    absorption.}
  \label{fig:multilambda3}
\end{figure}

In addition, we have the modes
\begin{gather}
  \label{eq:23}
  \hat b_q = \frac{1}{R + \frac{\Omega_Q}{g_Q}} \frac{\Omega_q}{g_q}
  \Bigl( \hat a_Q + \hat b_Q \Bigr) - \hat a_q, \qquad \qquad q \neq Q
\end{gather}
each coupling to the excited bright state $\ket{EB}$ and also to the
(absorbing) excited dark states $\ket{ED_q}$ (\fref{fig:multilambda3})
with strengths $g_q^{EB}$ and $g_q^{ED_r}$ (whose explicit form is not
of interest). These modes do not experience EIT.

\section{Propagating Fields}
\subsection{Dark-state polaritons}

The preceding section demonstrated that by a unitary transformation in
both the atomic states and the quantized field modes the
multi-$\Lambda$-system can be mapped to the well known standard
EIT-scheme. In order to apply the dark-state polariton formalism to
this system, we need to derive the wave propagation (Maxwell-Bloch)
equation for the field $\tilde b_Q$. For reference, we first rewrite
the main definitions of Ref.
\cite{fleischhauer02:_quantum_memory_for_photons} in our notation.

We introduce the atomic operators
\begin{equation}
  \label{eq:25}
  \tilde \sigma_{\alpha,\beta}^{(j)} = \ketbra[jj]{\alpha}{\beta} e^{i  \frac{\omega_{\alpha \beta}}{c} z_j}
\end{equation}
acting on the $j$-th atom located at position $z_j$, with
$\omega_{\alpha \beta}$ being the laser frequency.  Assuming that the
transition energies of the quantized fields are well separated, the
electric field can be decomposed into components each interacting only
with their respective transition:
\begin{equation}
  \label{eq:26}
  \hat E(z,t) = \sum \limits_{q=1}^Q \hat E_q(z,t), \qquad  \hat E_q \textrm{ coupling } \ket{B} \leftrightarrow \ket{A_q}.
\end{equation}

We now define the slowly varying field operators $\tilde a_q(z,t)$ by
the positive frequency parts of our field components:
\begin{equation}
 \label{eq:27}
   {{\hat E_q}^+}(z,t)  = \tilde a_q(z,t) \sqrt{\frac{\hbar \nu_q}{2 \varepsilon_0 V}} \enspace \exp\left[{i \frac{\nu_q}{c}(z-ct)}\right].
\end{equation}
To describe the evolution of the atomic variables, we can assume that
the quantum amplitude of the atomic variables does not depend strongly
on the position.  By introducing a ``smearing kernel''~$s$ with
$\int_0^L s(z) \,dz=\frac{L}{N}$ and a zero-centered support with a
width that is large compared to the average distance of two atoms but
small in relation to the medium length $L$, we obtain the mean-field
operators
\begin{equation}
  \label{eq:30}
  \tilde \sigma_{\alpha,\beta}(z) = \sum_{j=1}^N s(z-z_j)  \, \tilde \sigma_{\alpha,\beta}^{(j)},
\end{equation}
so that, assuming $\Delta=\delta=0$,
\begin{gather}
  \label{eq:31}
  \begin{split}
    \hat H(t) & = - \hbar \frac{N}{L} \int \limits_0^L \sum \limits_{q=1}^Q \Bigl[ \,g_q \, \tilde \sigma_{A_q,B}(z) \tilde a_q(z,t) \\
    & \quad + \tilde \sigma_{A_q,C}(z) \Omega_q(t) \Bigr] \,dz \qquad
    +\text{H.a.}
  \end{split}
\end{gather}
Performing mappings $\hat U$ and $\hat W$ on atoms and light, the
Hamiltonian transforms as follows:
\begin{gather}
  \label{eq:32}
  \begin{split}
    \hat H_\sub{int} & =  - \hbar \frac{N}{L} \int \limits_0^L dz \Biggl( \quad g \, \tilde \sigma_{EB,B} \, \tilde b_Q + \tilde \sigma_{EB,C} \, \Omega \\
    & +\sum \limits_{q=1}^{Q-1} \tilde b_q \Bigl( g_q^B \tilde
    \sigma_{EB,B} + \sum \limits_{r=1}^{Q-1} g_q^{ED_r} \tilde
    \sigma_{ED_r,B} \Bigr) + \text{H.a.} \Biggr).
  \end{split}
\end{gather}

The Maxwell-Bloch equations for the individual fields are
\begin{equation}
  \label{eq:33a}
  \Bigl( \frac{\partial}{\partial t} + c \frac{\partial}{\partial z} \Bigr) \tilde a_q = i N g_i^* \tilde \sigma_{B,A_Q}.
\end{equation}
Performing summation of eqs. (\ref{eq:33a}) over $q$'s with weights
$\Omega_q^*/g_q^*$ and utilizing relations (\ref{eq:12}) and
(\ref{eq:22}), we find
\begin{equation}
  \label{eq:33}
  \Bigl( \frac{\partial}{\partial t} + c \frac{\partial}{\partial z} \Bigr) \tilde b_Q = i N g \tilde \sigma_{B,EB}.
\end{equation}
In other words, if there is no light in the modes $\tilde b_{q\neq Q}$
and no atoms are in the excited states, the propagation of mode
$\tilde b_Q$ in a multilevel EIT setting is fully equivalent to that
in a single-$\Lambda$ system defined by \fref{fig:multilambda3}.

Similarly to ref.~\cite{fleischhauer02:_quantum_memory_for_photons},
one can define the dark-state polariton for this system. Upon entering
the medium an incoming light pulse in the EIT mode forms a
polariton~$\hat \Psi$, a superposition of an electromagnetic wave in
the $\tilde b_Q$ mode and a collective atomic excitation $\tilde
\sigma_{EB,C}$ which generates an eigenstate of eigenvalue zero of the
interaction Hamiltonian.

\begin{gather}
  \begin{split}
    \label{eq:34}
    \hat \Psi = \cos \theta(t) \, \tilde b_Q - \sin \theta(t) \sqrt{N}
    \, \tilde \sigma_{B,C}
  \end{split} \\
  \begin{split}
    \tan \theta(t) = \frac{\sqrt{N}}{R(t)}
  \end{split}
\end{gather}
By changing the classical control fields' parameter~$R$ the character
of this polariton (whether it is more optical ($\theta\approx 0$) or
has a stronger atomic component ($\theta \approx \tfrac{\pi}{2}$)) can
be changed.

\subsection{Incoupling and Slowdown} 
The susceptibility for the EIT mode \cite{scully97:_quant_optic} is
proportional to
\begin{equation}
  \label{eq:35}
  \chi_Q \propto  N g^2 \frac{\Delta - \delta}{(\Delta - \delta)(\delta + i \frac{\gamma}{2}) + \Omega^2}.
\end{equation} where
$\gamma$ is the spontaneous decay rate of the excited bright-state
$\ket{EB}$.  So for a signal beam in precise two-photon resonance
($\Delta=\delta$) the refractive index is one: no back-reflection or
absorption of a signal entering and passing through the medium occurs.
This also holds for pulses as long as their bandwidth is significantly
smaller than the EIT transparency window
\begin{equation}
  \label{eq:36}
  \textrm{FWHM} = \frac{\gamma}{2} \left(\sqrt{\left(\frac{4 \Omega}{\gamma}\right)^2+1}-1\right).
\end{equation}

If the effective Rabi frequency~$\Omega$ is small compared
to~$\gamma$, the transparency window is narrow and \fref{eq:35}
predicts a strong dispersion. This leads to a strongly reduced group
velocity $v_g$ for the polariton wave

\begin{equation}
  \label{eq:37}
  v_g = \frac{c}{1+n_g}, \quad  n_g=\frac{N}{R^2}.
\end{equation}

\section{RATOS}
\subsection{The procedure}
Based on this formalism we now describe a protocol for transfer of
quantum information between optical modes (Raman adiabatic transfer of
optical states, RATOS).

If the intensities of the control fields are changed slowly, the
eigenstates follow the new conditions adiabatically~\footnote{Strictly
  speaking, time variation of $W_{Qq}$s leads to a geometric phase
  which we will explore in a subsequent publication.}. The dark-state
polariton as eigenstate of zero interaction energy is thus preserved
-- however its mode composition and propagation velocity can be
controlled by the parameters $\{\Omega_q\}$ of the strong control
fields according to \fref{eq:22}.

This allows for transfer of quantum information from an optical mode
$\tilde a_i$ to another mode $\tilde a_j$:
\begin{itemize}
\item First only one strong control field~$\Omega_i$ is switched on.
  The medium then exhibits electromagnetically induced transparency
  for the $\tilde b_Q=\tilde a_i$ mode.
\item An incoming quantum pulse in the $\tilde a_i$ mode can enter the
  EIT medium without absorption or reflexion since at two-photon
  resonance the refractive index for the signal field is 1. The pulse
  experiences a reduction of the group velocity according to
  \fref{eq:37}.
\item This slowdown also leads to a spatial compression: the pulse
  gets shorter in length, which helps in keeping the size of the
  medium reasonably small.
\item Once the pulse is completely inside the medium, the control
  field~$\Omega_i$ is replaced by another field~$\Omega_j$
  adiabatically.  Assuming the mixing angle $\theta$ is kept constant,
  the polariton changes its characteristics as follows:
  \begin{equation}
    \begin{split}
      \label{eq:44}
      \hat \Psi_{t=-\infty} & = \cos \theta\, \tilde a_i - \sqrt{N} \sin{\theta} \, \tilde \sigma_{C,EB} \\
      & \to \cos \theta \, \tilde a_j - \sqrt{N} \sin{\theta} \,
      \tilde \sigma_{C,EB} = \hat \Psi_{t=\infty}
    \end{split}
  \end{equation}
  and all photons are now in the $j$-th mode.
\item A pulse with a different frequency but in the same optical
  quantum state as the original pulse exits the medium in mode $\tilde
  a_j$.
\end{itemize}

RATOS might find applications as an optical switch to route optical
quantum information. If in the end not one but several control fields
are present, the incoming pulse is split into optical modes with
different frequencies.

We now review a few recently published procedures for transferring
optical information via atomic transitions that are related to the one
developed above.  Zibrov et al.~\cite{zibrov02:_trans} used the
double~$\Lambda$ system formed by the fine structure splitting of
$^{87}$Rb. They first couple in a light pulse resonant to one of the
fine transition lines and store it via an adiabatic turn-off of the
control field.  Later on they retrieve it with a control field tuned
to the other fine structure transition.  Matsko {\it et al.} and Peng
{\it et al.}  \cite{matsko01:_nonad,peng05:_pulse} investigate
transferring a light pulse to another mode using storage in a
single-$\Lambda$ system. By interchanging the roles of the control and
signal modes in the retrieving process, the pulse is retrieved at the
frequency of the original control field.  The main difference of RATOS
with respect to these proposals is that it offers a way to extend this
transfer to multiple modes (and even to their coherent superpositions)
and that an intermediate storing of the pulse is not necessary.

\section{Quantum state engineering}
Now we also can answer the question to which extent RATOS can be used
for engineering of optical quantum states. The only mode that can
losslessly enter a multilevel EIT sample is that associated with the
operator $\hat b_Q$ which is a linear combination of individual mode
operators $\{\hat a_q\}$. However, by choosing amplitudes and phases
of the control fields one can adjust the coefficients of the linear
combination.

\begin{figure}[tbp]
  \begin{center}
    \includegraphics[width=\columnwidth,keepaspectratio]{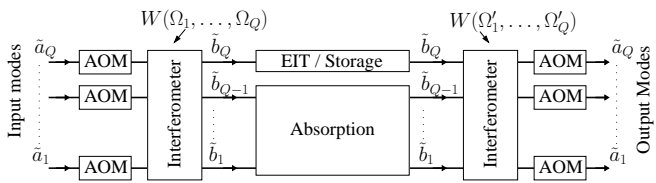}
  \end{center}
  \caption{Linear optical circuit equivalent to a multi-$\Lambda$ configuration. The phase shifts and reflectivities of the input (output)
    combining mirrors are determined by the phases and amplitudes of
    the classical control fields during the in-coupling (out-coupling)
    process. In this model, the acousto-optical modulators (AOM) at
    cell entrances and exits bring the input fields to the same
    frequency so they become indistinguishable when handled by the
    interferometers.}
  \label{fig:equiv}
\end{figure}

The linear transformation $W(\{\Omega_1,\ldots,\Omega_Q\}):\ \tilde
a_q\to \tilde b_q$ of the fields at the cell entrance can be
visualized as an interferometer, i.e. a sequence of linear optical
elements such as beam splitters and mirrors (\fref{fig:equiv}).  While
this transformation does not by itself represent any physical process,
the modes $\tilde b_q$ do have a physical meaning as only one of them
is able to propagate through the cell due to EIT; the rest get
absorbed.  While the mode $\tilde b_Q$ is traversing the cell, the
control fields may change adiabatically so at the cell output, when
the propagating modes convert back to $\tilde a_q$'s, the
interferometer, defined by $W(\{\Omega'_1,\ldots,\Omega'_Q\})$, may be
completely different.  As a result, optical states can be transferred
among different input and output modes~$\tilde a_q$.

As an example, we consider a double-$\Lambda$-system ($Q=2$) with two
control fields such that $\tfrac{\Omega_1}{g_1} =
\tfrac{\Omega_2}{g_2}$. Then the incoming light fields can be
decomposed into the orthogonal modes $\tilde b_{1/2} =
\frac{1}{\sqrt{2}} \left(\tilde a_2 \mp \tilde a_1 \right)$. Light in
the mode $\tilde b_Q=\tilde b_2$ sees EIT and is subject to the RATOS
process. Light in the mode $\tilde b_1$ however couples to both
excited states $\ket{ED_1}$ and $\ket{EB}$. This leads to absorption;
due to spontaneous emission excitations of this mode will be scattered
away or decay into the EIT mode. This agrees with Raczy\'nsky's and
Zaremba's predictions for a classical double-$\Lambda$-system
\cite{raczynski04:_elect}.

Suppose this system is irradiated by an optical pulse which contains
exactly one photon in each mode. The optical mode associated with this
pulse consists to equal parts of the EIT-mode~$\tilde b_Q=\tilde b_2$
and the absorbing mode~$\tilde b_1$.
\begin{equation}
  \label{eq:45}
  \tilde a_1^\dagger \tilde a_2^\dagger \ket{0} = \frac{1}{2} \left( {\tilde b_2^\dagger}\,^2 - {\tilde b_1^\dagger}\,^2 \right) \ket{0}.
\end{equation}
For this reason, only with $50\%$ probability will both photons be
coupled into the medium and get fused into the EIT mode $\tilde b_2$;
with equal probability they will experience absorption. So in this
setup the double-$\Lambda$ medium does not perform better than an
ordinary beam splitter: here the Hong-Ou-Mandel effect
\cite{hong87:_measur} also provides a $50\%$ probability for the two
photons to fuse into a specific mode. Furthermore, it is clear that no
combination of control fields would make the atomic system fully
transparent for the state (\ref{eq:45}), so this state cannot be
coupled into the EIT medium without loss.

In summary, a multi-$\Lambda$ medium is equivalent to a linear optical
system with a built-in storage device and with multiple input and
output modes which differ in frequency (\fref{fig:equiv}).

\section{conclusions}
We have extended a full quantum treatment of the
electromagnetically-induced transparency to multi-$\Lambda$ systems.
An explicit form of an unitary mapping is presented that relates the
dark states to the effects observed in a standard EIT scheme. Most of
the properties of this well investigated system can be transferred and
extended to systems with multiple excited levels.

The mapping provides a physical explanation for the existence of the
decay sensitive $\ket{ED_q}$ states and the according bright-state
modes $\hat b_{q\neq Q}$.

EIT in a multi-$\Lambda$ scheme might be useful for multiplexing and
routing of optical quantum information as well as for the preparation
of multi-mode entangled quantum states. Its application to
quantum-optical engineering is however limited by its equivalence to a
linear-optical setup with a built-in storage capability.

We thank M.~Fleischhauer, B.~Brezger, A.~Raczy\'nsky, J.~Zaremba, and
B.~Sanders for helpful discussions. This work was supported by NSERC,
CIAR, and CFI.

\bibliography{multilambda}

\end{document}